\documentclass[twocolumn]{aastex631}

\newcommand{\msun}{M$_{\odot}$}
\usepackage[mathscr]{euscript}

\begin{document}

\title{Intermediate Mass Ratio Inspirals in Milky Way Galaxies}

\correspondingauthor{Jillian Bellovary}
\email{jbellovary@amnh.org}

\author[0000-0001-7596-8372]{Jillian Bellovary}
\affiliation{Department of Physics, Queensborough Community College, 222-05 56th Ave, Bayside, NY 11364, USA}
\affiliation{Astrophysics Program, CUNY Graduate Center, 365 5th Ave, New York, NY 10016, USA}
\affiliation{Department of Astrophysics, American Museum of Natural History, New York, NY 10024, USA}

\author{Yuantong Luo}
\affiliation{Department of Physics, Queensborough Community College, 222-05 56th Ave, Bayside, NY 11364, USA}

\author{Thomas R. Quinn}
\affiliation{Astronomy Department, University of Washington, Seattle, WA 98195, USA}

\author{Ferah Munshi}
\affiliation{Physics and Astronomy Department, George Mason University, 4400 University Drive, MSN 2A1, Fairfax, VA 22030, USA}

\author{Michael Tremmel}
\affiliation{School of Physics, University College Cork,   College Road, Cork T12 K8AF, Ireland}

\author{James Wadsley}
\affiliation{Physics and Astronomy Department, McMaster University, 1280 Main Street West ,Hamilton, Ontario, L8S 4K1, Canada}

\begin{abstract}

A consequence of a non-zero occupation fraction of massive black holes (MBHs) in dwarf galaxies is that these MBHs can become residents of larger galaxy halos via hierarchical merging and tidal stripping.  Depending on the parameters of their orbits and original hosts, some of these MBHs will merge with the central supermassive black hole in the larger galaxy.  We examine four cosmological zoom-in simulations of Milky Way-like galaxies to study the demographics of the black hole mergers which originate from dwarf galaxies.  Approximately half of these mergers have mass ratios less than  0.04, which  we categorize as intermediate mass ratio inspirals, or IMRIs.  Inspiral durations range from 0.5 - 8 Gyr, depending on the compactness of the dwarf galaxy.  Approximately half of the inspirals may become more circular with time, while the eccentricity of the remainder does not evolve.  Overall, IMRIs in Milky Way-like galaxies are a significant class of black hole merger that can be detected by LISA, and must be prioritized for waveform modeling.

\end{abstract}


\section{Introduction} \label{sec:intro}

During the past decade or so, the existence of massive black holes (MBHs) in dwarf galaxies has been demonstrated in multiple ways.  Evidence for these objects has been found in X-rays \citep{Lemons15,Baldassare17,Mezcua18,Birchall20}, optical emission lines \citep{Reines13,Moran14,Molina21,Polimera22}, radio \citep{Mezcua19,Reines20}, infrared \citep{Satyapal14}, masers \citep{Zaw20}, and variability \citep{Baldassare20,Lira20}.  The fraction of dwarfs which host MBHs remains uncertain \citep[see][for a thorough review]{Greene20} but is certainly significant at larger dwarf masses ($\gtrsim 10^9$ M$_*$).  The formation mechanism of these MBHs is unknown, but they broadly seem to follow the known galaxy-M$_{\rm BH}$ scaling relations and have ``intermediate'' masses (e.g. $10^3 - 10^5$ \msun).  These masses indicate that there could be a seed stage of MBH growth in the early universe, where black holes form within the $10^3 - 10^5$ \msun~ mass range and may then grow to reach supermassive size extremely rapidly.  If these MBHs in dwarfs are ``failed'' supermassive black holes, they provide a unique look into black hole formation at the earliest epochs of the universe.  

Of course, dwarf galaxies have cosmic significance besides hosting MBHs.  Dwarfs are plentiful in the environments of more massive galaxies, and regularly merge with them.  This process of hierarchical merging is known to build up the stellar halo of massive galaxies \citep{Zolotov09,Font11,Tissera12,Deason16,Fattahi20} and create stellar streams in galactic halos \citep{Helmi99,Yanny03,Shipp18,FIREstreams}.  This process is occurring in our own Milky Way with galaxies such as the Sagittarius Dwarf \citep{Ibata97,Majewski03,Belokurov06}.  The nearby Magellanic Clouds are on a collision course with the Milky Way as well, and are generally assumed to be on their first passage through the Galaxy \citep{Besla07,Kallivayalil13,Jethwa16} (however see  \citet{Vasiliev23}).  Our neighboring disk galaxy  M31  also shows evidence of interaction with its dwarf neighbors.    The dwarf elliptical galaxies NGC205 and NGC 147 both show evidence of tidal features \citep{Choi02,Johnston02,Ferguson02,Crnojevic14}, and the Great Stellar Stream may be a result of a close passage of the low-mass disk galaxy M33 2 Gyr ago \citep{Bernard12,Bernard15a}.  Numerous other streams around M31 have unidentified hosts but are likely due to disrupted dwarf galaxies \citep[][and references therein]{Ferguson16}.  The evolution of massive disk galaxies such as the Milky Way and M31 is clearly shaped by the accretion and tidal stripping of neighboring dwarfs.


Eventually  merging dwarfs are tidally disrupted, and their contents  become indistinguishable from the main galaxy.  If a dwarf galaxy hosts an MBH, the black hole will also join the remnants of its host in the halo.  Simulations have predicted wandering MBHs from disrupted dwarfs in galaxy halos for quite some time \citep{Bellovary10,Tremmel18,Weller22}.   Depending on a number of factors which govern the dynamical timescales of these objects, these wandering MBHs may remain in the halo for more than a Hubble time, or their orbits may decay and they can merge with the central supermassive black hole (SMBH).  For example, in instances where the host dwarf resists tidal stripping and retains its mass as it inspirals, it can deliver the MBH to the SMBH quickly due to increased dynamical friction.  Such mergers may be a prominent growth mechanism for supermassive black holes, especially at the lower-mass end of the SMBH population.  In addition, these mergers will produce gravitational waves, which could be detectable by the upcoming LISA (Laser Interferometer Space Antenna) mission.

LISA is a  gravitational wave detector which will launch in the mid-2030s and operate for 4-10 years \citep{LISA,LISA_Redbook24}.  Its 2.5 million km baseline makes it able to detect mergers of black holes in the intermediate mass range to redshift $z = 20$ and beyond, including phenomena such as Milky Way-like  SMBHs consuming intermediate mass black holes (IMBHs) delivered by merging dwarf galaxies.  This type of merger is likely to have a mass ratio of $\lesssim 0.01$, which is known as an Intermediate Mass Ratio Inspiral, or IMRI (defined as $10^{-2} > q > 10^{-5}$) \citep{LISAWhitePaper}.  More specifically it is known as a ``heavy IMRI,'' which is an SMBH - IMBH merger.   Gravitational wave detections of IMRIs have the unique capability to determine black hole masses within a few percent; observations such as these will allow tight constraints on dwarf MBH masses and possibly provide limits for seed formation mechanisms as well.  However, progress is needed in waveform modeling to allow for the proper interpretation of gravitational wave signals from these IMRI events.  See section 3.2 of \citet{LISAWhitePaper} for more details on IMRIs of all types.  

Motivated by the need to model IMRI waveforms for the upcoming LISA mission, we present here a census of SMBH-IMBH mergers from zoom-in cosmological simulations.  Knowledge about the mass, redshift, and eccentricity distribution of these events will inform further studies and advance our preparation for the LISA mission.  In Section \ref{sec:methods} we detail the simulations, including the relevant black hole physics. In Section \ref{sec:results} we discuss the basic demographics of the IMRI events, and in Section \ref{sec:evol} we present results on how the inspirals evolve with time.  In Sections \ref{sec:LISA} and \ref{sec:concl} we discuss the repercussions of our results for the LISA mission and provide an overall summary.

\section{Simulations} \label{sec:methods}

We use the simulation suite known as   the DC Justice League, individually named Sandra, Ruth, Sonia and Elena.  These four simulations each consist of a Milky Way-like galaxy with surrounding cosmological environment run to $z = 0$, and are described in detail in \citet{Bellovary19}.  We summarize their properties below. 

 \subsection{The DC Justice League}
 
 The DC Justice League simulations were created with  ChaNGa, an $N$-Body Tree + smoothed particle hydrodynamics (SPH) code \citep{Menon15} which employs dynamic load balancing via the Charm++ framework, allowing for improved scalability up to 100,000+ cores.  The SPH kernel is calculated using a geometric mean density in the SPH force expression \citep{Ritchie01, Menon15,Wadsley17} which accurately represents Kelvin-Helmholtz instabilities and other contact discontinuities.   The initial conditions were selected from a 50 Mpc volume using Planck cosmological parameters \citep{Planck16}.  Using the volume renormalization technique from \citet{Katz93} we created ``zoom-in'' versions of Milky Way-like galaxies (similar in terms of mass and morphology) to resimulate at high resolution.  
 
 Each run is at ``near mint'' resolution, defined as a force softening resolution of 170 pc, dark matter particle masses of $4.2 \times 10^4$ \msun, gas particle masses of $2.7 \times 10^4$ \msun, and star particle masses of 8000 \msun.  We model a UV background based on \citet{Haardt12}, use a molecular hydrogen-based star formation prescription \citep{Christensen12}, blastwave supernova feedback \citep{Stinson06}, and metal diffusion and cooling \citep{Shen10}.   To identify individual galaxies, we use the Amiga Halo Finder \citep{Knollmann09} which uses an overdensity criterion for a flat universe to identify individual halos \citep{Gill04}.  
 
The DC Justice League simulations have been studied extensively and been found to illuminate several characteristics of dwarf galaxies in Milky Way-like galaxy environments.  \citet{Christensen24} found differences in dwarf galaxy properties based on whether they form near a Milky Way-mass galaxy or not, and  \citet{Akins21} explored the role of star formation quenching in dwarfs as they infall into the larger halo.    Additionally,  \citet{Munshi21} included them in a study of the stellar mass - halo mass relation for the lowest mass galaxies, which is highly dependent on environment. This simulation suite is an excellent tool to probe the repercussions of dwarfs infalling into more massive galaxies.

\subsection{Black Hole Physics}
 
We use black hole  formation and evolution prescriptions that are unique to ChaNGa and the DC Justice League, and describe them here.  Black hole (BH) particles form based on the properties of their parent gas particle, with no dependence on global halo properties.  To form a BH a gas particle must meet several criteria, which broadly mimic those of the direct collapse formation mechanism \citep[e.g.][]{Begelman06, Volonteri12}.  Specifically, gas particles must be dense ($\rho > 3000$ cm$^{-3}$), have low metallicity ($Z < 10^{-4}$), low molecular hydrogen fraction ($f_{H2} < 10^{-4}$), and a temperature less than $T < 2 \times 10^4$ K.  In addition, the gas must meet a Jeans mass criterion  $M_{\rm Jeans} =  (\pi^{5/2}c_s^2) / (6\rho^{1/2}) > 4 \times 10^5$\msun.  These criteria ensure that the parent gas particle is in a region that is likely to collapse, cool via atomic hydrogen only,  could potentially form a direct-collapse black hole seed.  The mass of each  BH seed is 50,000 \msun.

BH particles accrete gas based on a modified version of the Bondi-Hoyle formula, which adjusts the accretion rate based on the local gas density and angular momentum, and is described in detail in \citet{Tremmel17} and \citet{Bellovary19}.  BHs impart thermal feedback energy from accretion isotropically into the surrounding gas based on a prescription proportional to the accretion rate, where we assume a radiative efficiency $\epsilon_r$ = 0.1 and a feedback coupling efficiency of $\epsilon_f$ = 0.02.  As reported in \citet{Bellovary19} and \citet{Bellovary21}, the accretion rates of BHs in the lower-mass galaxy environments we study here are quite low, and we do not expect our choice of subgrid models and parameters to have a strong affect on any of our results.

Our prescriptions for dynamical friction and BH mergers are highly relevant to our work here, and we describe them in full detail.  We implement a subgrid dynamical friction model based on the Chandrasekhar formula \citep{Chandrasekhar43,Binney08} which is described in \citet{Tremmel15}.  This prescription is vital because the BH particle does not feel realistic dynamical friction in the simulation, since the surrounding particles are of comparable mass.  Our model mimics the BH moving through a sea of smaller objects by including an additional acceleration as follows:

\begin{equation}
{\bf a}_{\rm DF} = -4\pi G^2 M_{\rm BH} \rho (< {\bf v}_{\rm BH})  \rm{ln} \Lambda  \frac{{\bf v}_{\rm BH}}{v_{\rm BH}^3}
\end{equation}

where $\rho (< {\bf v}_{\rm BH})$ is the density of collisionless particles moving slower than the BH, $M_{\rm BH}$ and  $v_{\rm BH}$ are the mass and velocity of the BH, respectively, and ln $\Lambda$ is the Coulomb logarithm.  This latter quantity depends on the impact parameters $b_{min}$ and $b_{max}$, such that ln $\Lambda \sim$ ln ($\frac{b_{max}}{b_{min}}$).  We set $b_{max}$ equal to the softening length, because dynamical friction is well-resolved on scales this size and larger.  For the minimum impact parameter, we set it to be the minimum deflection radius, $b_{min} = GM_{BH} / v_{\rm BH}^2$, with a minimum possible value of the Schwarzschild Radius.  The resulting acceleration is added to the BH's current acceleration and integrated at every timestep, such that dynamical effects from structures smaller than our resolution element can be approximately accounted for.  We note that while the BH within a galaxy is subject to this dynamical friction model, dynamical friction BH's host dwarf galaxy inspiralling into a larger host is captured by the simulation and modeled correctly via the $N$-body method.

Our simulations do not resolve the full inspiral process of a BH pair; instead, we merge BH particles together when the equivalent of a close pair has formed.   BH particles merge when they come within less than two softening lengths and also meet the criterion $\frac{1}{2}\Delta {\vec{ \rm v}} < \Delta {\vec{ \rm a}} \cdot \Delta {\vec{ \rm r}}$, where $\Delta {\vec{ \rm v}},  \Delta {\vec{ \rm a}}$ and  $\Delta {\vec{ \rm r}}$ represent the relative velocity, acceleration, and radius vectors of the two BHs respectively.  This latter criterion ensures that the BHs have low relative velocities and prevent mergers during e.g. rapid flybys.  When the BHs merge, their masses are summed and the remnant is positioned at the center of mass of the system.  This merger methodology is a crude and fairly inaccurate representation of the actual physics of black hole mergers, which include a gravitational recoil kick as well as mass loss in the system from radiated gravitational wave energy.  Recoil kicks can range from $10 - 1000$ km s$^{-1}$  which is often enough to eject a remnant black hole out of its host galaxy at high redshifts, when escape velocities are small.  The effect of gravitational recoil in cosmological simulations has been studied in a post-processing capacity by \citet{Dunn20} and on-the-fly by \citet{DongPaez24}, and both find that massive black hole growth is stifled when black holes are ejected from galaxies and no longer contribute to the black hole mass function.    However, for IMRI events with low mass ratios, both the mass loss and the recoil kick are predicted to be smaller \citep{Campanelli07,Baker08}.  IMRI recoil kicks of  $\sim 10$km s$^{-1}$  would have little effect on disturbing the remnant's position from the center of the host galaxy.

\subsection{Limitations of the Model}\label{sec:limits}

Because black hole seeding occurs relatively stochastically, occasionally multiple BHs form within the same region at the same time.  In these instances, they often merge together immediately.  We do not count these as bona fide mergers, but rather a result of the formation process.  Our initial mass BH of 50,000~\msun~ is somewhat arbitrary, and a factor of two or three is certainly within the realm of possible masses of direct collapse seeds.  This process is effectively a slightly varying initial mass function, which while limited from below by our resolution captures the possibility of larger mass regions being susceptible to direct collapse.  The initial masses of each BH in our simulations are presented in Figure 1 of \citet{Bellovary19}.   The majority range from $4.5 <  \log(M_{BH}/M_\odot)  < 6$ \msun, with a handful reaching $10^7$ \msun.  

This variation in seed mass has an effect on our estimates for merger mass ratios.  Since either black hole mass in a merger has a numerical uncertainty of a factor of 2 or 3 due to early merging, mass ratios could be skewed upwards or downwards by this amount, or if both black holes are affected by early mergers the effect could approximately cancel out.  Since the actual initial mass function of black hole seeds is unknown, no mass ratio can be precisely accurate.  The focus of this work is on intermediate mass ratios, which are an order of magnitude different from what is classified as a major merger.  This difference creates a gap in parameter space where it is moderately straightforward to classify IMRIs as different from more major black hole merger events, regardless of the uncertainty in the exact mass ratio.

Another limitation of our model, which is due to resolution, is that the BH merger methodology described above omits a great deal of physics.  In our model, the final few $\sim 100$ pc of simulated binary coalescence occurs instantaneously, but in reality several processes govern the actual timescale.  For example, dynamical friction due to background matter acts on scales of 1 - 100 pc.  At closer separations, loss-cone scattering of stars in the central region shrinks the separation of black holes,  as does emission of gravitational radiation.  The presence of gas can also affect the coalescence of binary black holes at all scales.  Each of these processes has its own timescale on which it acts, depending on the environment of the binary, but none of them are so quick to make the timescale negligible.   To highlight the importance of this multitude of processes, in their absence it would take $\sim 10^{30}$ years for gravitational emission alone to bring black holes from the scale of our simulation resolution to merger (see \citet{Chan18} equation 4), underscoring the need for further simulations to resolve and document these timescales more accurately.

Estimates of the binary hardening time-scale in massive galaxies range from  $10^7-10^8$ years \citep{Armitage02,Haiman09, Colpi14, Holley-Bockelmann15}, though full coalescence timescales from large distances can be much longer depending on mass ratio and total system mass  \citep{Katz2019,Volonteri20}.   This timescale can become even longer in less massive galaxies, due to the nature of the shallower potential wells of dwarf galaxies.  \citet{DeCun23} showed that BH mergers in dwarf galaxies can take up to a few $10^9$ years due to inefficient dynamical friction in galaxies with cored density profiles (though some estimates for low-mass galaxies are shorter \citep{Khan2021} and some longer  \citep{Tamfal18}).  In this work we present the inspiral time of mergers of IMBHs with SMBHs throughout cosmic time, often in low-mass galaxies, but our simulations do not take the abovementioned delays into account.  We will discuss the repercussions of this issue in the following sections.

\section{Demographics of the Mergers} \label{sec:results}

For this work, we follow the  merger tree of the $z = 0$ central black hole of each Milky Way-like galaxy and analyze all black hole mergers within the tree.   In each case the central black hole is also the most massive black hole in the halo.  We trace the merger tree of each black hole back to its origins, finding all ``victims'' throughout cosmic time (omitting mergers immediately after BH particle formation, see Section \ref{sec:limits}).  Since galaxy formation is hierarchical, some of these mergers occur not in the Milky Way-like galaxies themselves but in progenitors which exist as low-mass galaxies in the early universe.  We include the entire progenitor history of each black hole to create  a census of mergers which is as complete as possible.

\begin{figure}
\includegraphics[width=3.3in]{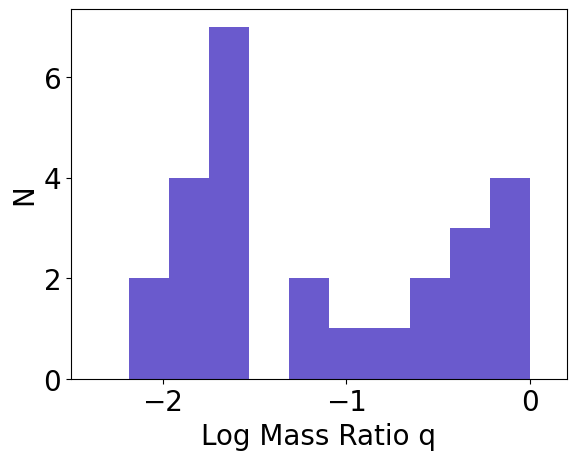}
\caption{Distribution of the log of merger mass ratios.  The distribution is somewhat bimodal, with a substantial fraction of mergers exhibiting log mass ratios between -2.2 and -1.5, or about 1:150 $< q <  $1:30.  
\label{fig:q}
}
\end{figure}

The distribution of black hole merger mass ratios is quite skewed towards small ($< 0.1$) values.  Figure \ref{fig:q} shows a histogram of the log of the mass ratios, which has an approximately bimodal appearance.  There is a prevalence of ratios with values    $ -2.2  < \log q < -1.5$  or 1:150 $< q <  $1:30, indicating a predominance of mergers in the IMRI regime.  The remainder of the mergers span the 1:10 $< q <$ 1:1 space, representing more equal-mass mergers.  We use this bimodality to delineate IMRIs from more equal-mass mergers, and henceforth color IMRIs as yellow in subsequent figures.  While the break in the distribution occurs at a 1:25 ratio or $q \sim 0.04$, rather than $q < 0.01 $ as defined by the community, we argue that this definition is reasonable because (a) our mass ratios are uncertain to a factor of a few (Section \ref{sec:limits}) and (b) the computational challenges of simulating IMRI waveforms extend to all $q$ values less than 0.1 (Section \ref{sec:LISA}).

\begin{figure}
\includegraphics[width=3.3in]{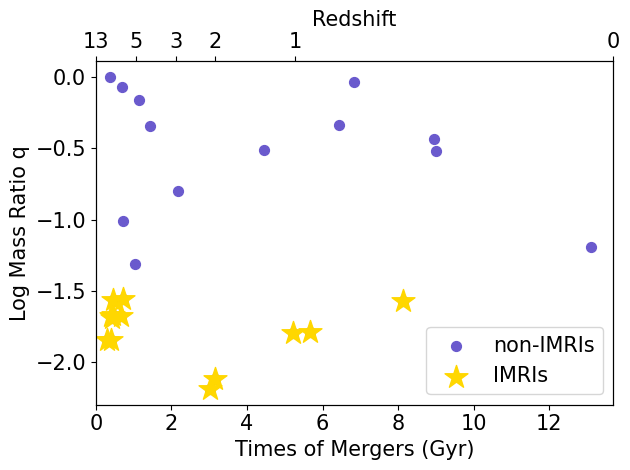}
\caption{ Log of mass ratio vs  merger times in Gyr since the big bang.   The purple circles represent non-IMRI mergers, while yellow stars are the IMRI mergers.  Both subsets  show similar activity with time, with increased mergers at the earliest epochs.
\label{fig:times}
}
\end{figure}

The majority of the mergers take place within the early universe -- about 3 Gyr after the Big Bang.  Figure \ref{fig:times} shows  the log of the mass ratio vs time of  merger, exhibiting the highest merger rates at the earliest times.  The purple circles show all  non-IMRI black hole mergers, while the yellow  stars are the IMRI mergers (identified from the left peak of Figure \ref{fig:q}).  These times do not account for unresolved delays in BH coalescence (see Section \ref{sec:limits}).   We expect these delays should be on the order of a maximum of 1 Gyr for close to equal-mass mergers in low mass galaxies  \citep{DeCun23}, but it is possible it could be longer for IMRI-type mergers. 

\begin{figure*}
\includegraphics[width=0.24\textwidth]{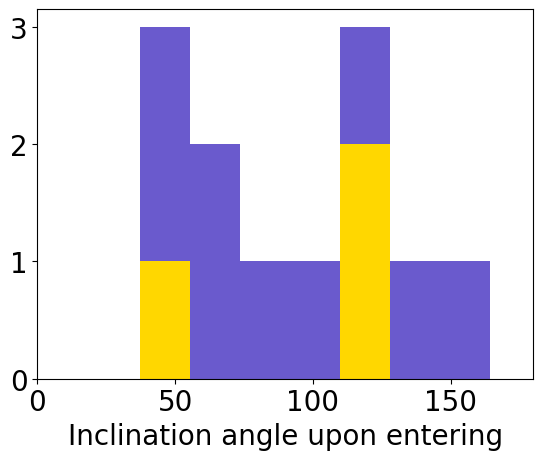}
\includegraphics[width=0.24\textwidth]{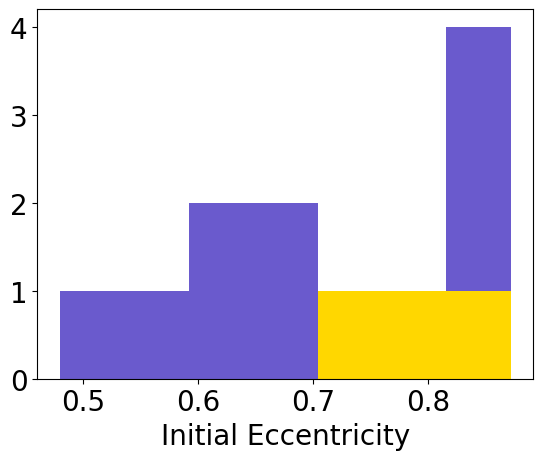}
\includegraphics[width=0.24\textwidth]{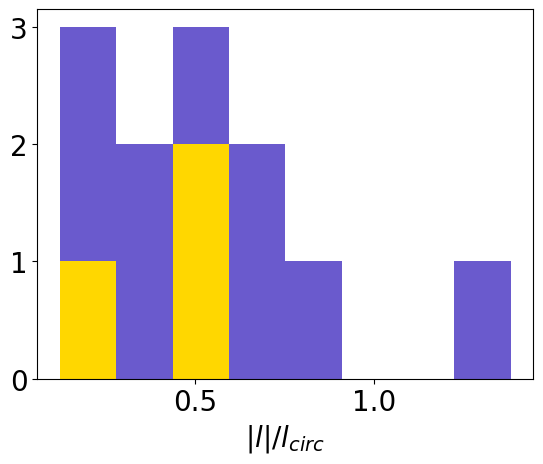}
\includegraphics[width=0.24\textwidth]{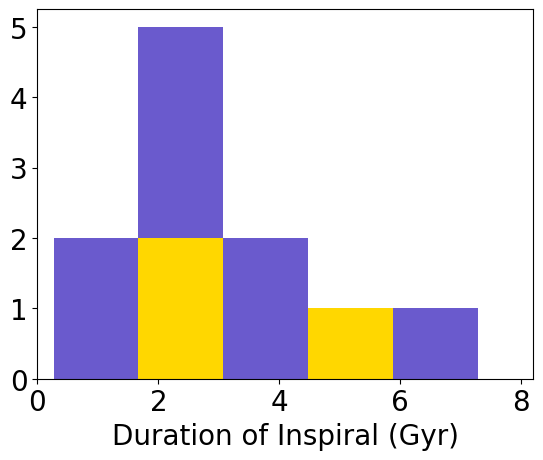}
\caption{Distribution of inclinations (far left), eccentricities  (center left),  angular momentum (center right), and merger durations in Gyr (far right).  The purple histograms represent all mergers, while histograms overplotted in yellow denote IMRIs.  This color scheme is used throughout the paper.
\label{fig:i_and_e}
}
\end{figure*}

The initial orbital parameters of each MBH is dependent on how its host galaxy enters the main halo, and this entry governs the overall evolution of much of the orbital inspiral and merger.  To investigate these details, we measure the initial inclination, angular momentum, and eccentricity of the entry of each MBH into the main halo.  The time of halo entry is determined by the first simulation snapshot where the BH is within the virial radius of the main halo.  At this moment the host dwarf galaxy may still be identified as an independent galaxy, before its tidal disruption is complete.   The resolution of this process is dependent on the time resolution of the snapshots, which is on average about 50 Myr.  Because the earliest BH mergers happen very quickly in small galaxies,  their halo entry and inspiral are not resolved and a full analysis is not possible.  Therefore, we are forced to omit some of the earliest mergers in our subsequent results for lack of time resolution.

We calculate the initial inclination by first measuring the position and velocity ($\vec r$ and $\vec v$) of the incoming BH relative to the center of the main halo.  We align the host galaxy with respect to the angular momentum of its gas disk, which gives us the angular momentum vector $\vec L$ for the main halo.  We calculate the angular momentum vector for the BH, $\vec l = \vec r \times \vec v$, and then use this in conjunction with the galaxy's angular momentum to calculate the inclination:

\begin{equation}
{\rm cos } ~i =  \frac{\vec l  \cdot  \vec L}{||\vec  l|| \hspace{4pt}||\vec L  ||}
\end{equation}

In this frame, $0^{\circ}$ represents an entry parallel to and into the disk, $90^{\circ}$ is a polar entry with respect to the disk, and $180^{\circ}$ is a retrograde disk-oriented entry.   In the far left panel of Figure \ref{fig:i_and_e} we show the distribution of the inclination angles of entry for each BH we are able to measure.   The distribution is consistent with isotropic, with an approximately equal number of prograde and retrograde entries.

\begin{figure*}
\includegraphics[width=0.5\textwidth]{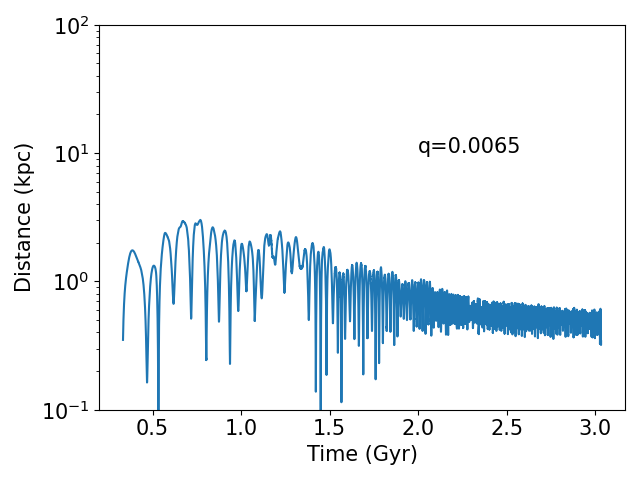}
\includegraphics[width=0.5\textwidth]{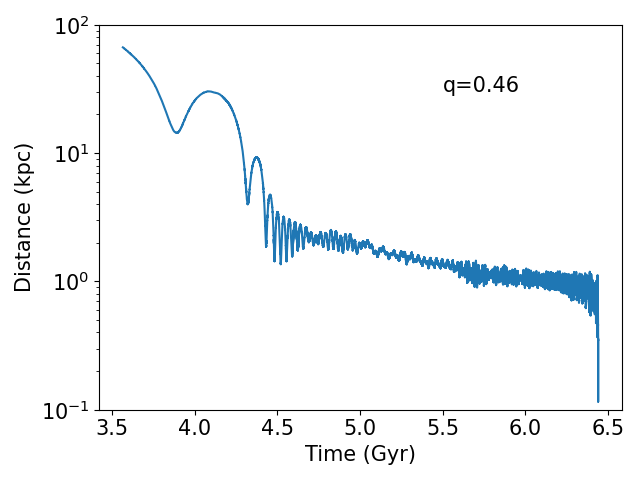}
\includegraphics[width=0.5\textwidth]{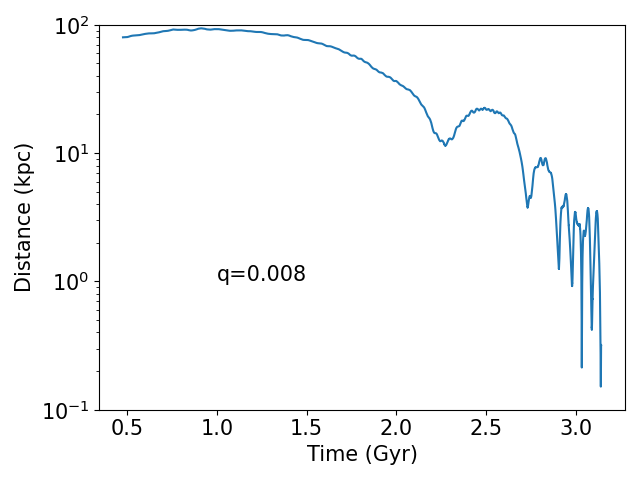}
\includegraphics[width=0.5\textwidth]{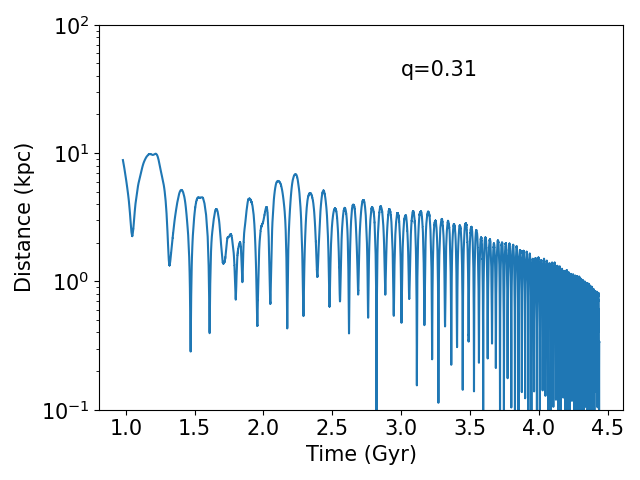}
\caption{The distance to the central supermassive black hole vs time (since the Big Bang) for four representative inspirals.     Mass ratios of each merger are noted on the panels.  Each plot begins at the time the BH enters the parent halo. 
\label{fig:inspirals}
}
\end{figure*}

We also measure the initial eccentricity of BHs as they enter the halo, shown in the center right panel of Figure \ref{fig:i_and_e}.  We calculate eccentricity by tracing the orbit of each BH after it crosses the virial radius of the main halo, and measuring its pericenter and apocenter distances.  Some orbital inspiral examples are shown in Figure \ref{fig:inspirals}, which shows the infall of an example black hole over time.  
We calculate eccentricity using the following formula:

\begin{equation}
e =  (r_{\rm apo} - r_{\rm peri}) / (r_{\rm apo} + r_{\rm peri})
\end{equation}

The initial eccentricity distribution is  fairly evenly distributed, with a slight preference for more radial orbits upon entry (Figure \ref{fig:i_and_e}, center right panel).  This distribution reflects the satellite merger infall distribution, which can also range from fairly circular to very eccentric.  We examine the evolution of eccentricity with time in Section \ref{sec:evol}.

Related to eccentricity, we show the distribution of incoming angular momenta in the center left panel of Figure \ref{fig:i_and_e}.  We calculate $|l|$ as the magnitude of the angular momentum the black hole at the entry point, and $l_{circ}$ as calculated angular momentum of a circular orbit at the same point.  We note there is one orbit where $|l| > l_{circ}$, which should not inspiral if energy is conserved.  However, since the orbit decays due to dynamical friction, this object's orbit can decay and the black hole can merge.


We estimate the duration of each merger by marking the snapshot where the BH enters the main halo as the starting point, and the moment when the two black holes coalesce in the simulation as the end.   The distribution of merger durations is shown in the far right panel of Figure \ref{fig:i_and_e}.  This process does not take into account unresolved dynamical friction or hardening (see Section \ref{sec:limits}), so these times should be taken as a lower limit.  Even so, the entire inspiral process is often a few to several billion years long. 

In Figure \ref{fig:inspirals} we show four examples of inspiraling BH trajectories with quite different properties.   The two left panels show examples of  IMRIs, with  mass ratios of $q$ =  0.0065 and 0.008.      The inspiral in the bottom left panel shows an example of where the infalling BH is not in the center of its host halo.  The small-scale oscillations represent an orbit within the host dwarf galaxy, and disappear once the dwarf disrupts and the BH reaches the center of the main galaxy.   In the right panels we see more equal mass inspirals ($q = 0.46$ and 0.31).  The upper right plot is a canonical inspiral, with characteristically decreasing apo- and peri-centers.  The BHs become quite close by 5 Gyr, but they do not meet the parameters for coalescence until 1.5 Gyr after that point.   

 To investigate trends in the orbital evolution of each BH based on the initial trajectory, we examine the relation of several quantities to the duration of the inspiral.  We place uncertainties on the inspiral duration as follows.  We are confident that our simulations physically capture the large-scale inspiral process, but when the black hole particles reach the resolution limit they may frequently 
 pass within two softening lengths of each other at much earlier times than they actually merge. 
 In some cases this interval is exceedingly long (several Gyr, see Figure  \ref{fig:inspirals} upper right).  Since this portion of the inspiral occurs at the resolution limit of the simulation, it is possible that these prolonged inspiral times are artificial.  We place the left end of the error bar at the moment when BHs reach the required distance for merging but do not actually do so.  The right end of each errorbar is 1 Gyr, representing a reasonable delay time for a black hole pair to coalesce after they reach a $\sim170$ parsec distance \citep[see][]{Katz2019,DeCun23}.
 
  
  
\begin{figure}
\includegraphics[width=0.45\textwidth]{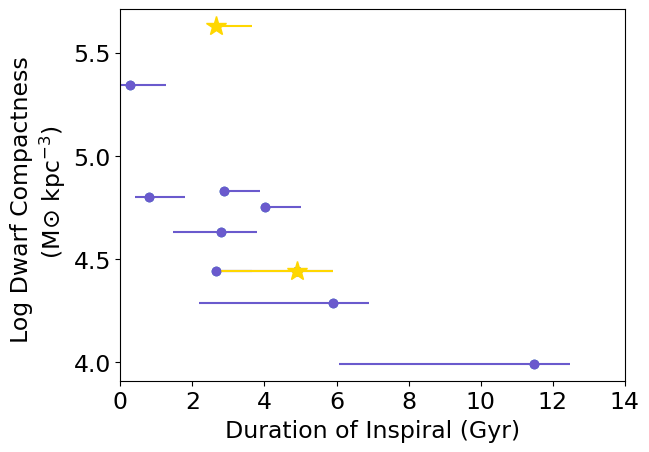}
\caption{ 
Log of the Compactness parameter vs inspiral duration in Gyr.   More compact galaxies have shorter inspiral times compared to those which  are less compact.   Points represent the time from halo entry until merger in the simulation.  Error bars are as described in the text, and yellow stars represent IMRI inspirals.
 \label{fig:durations}
}
\end{figure}

 The primary quantity which correlates with inspiral time is the compactness of the host galaxy.
 We devise a compactness  parameter as a general proxy for density, and calculate it by finding the half-mass radius of the stellar population of each host galaxy, and  dividing  the stellar mass within that radius by the cube of that radius at the time of infall.  At this moment the dwarf host has already undergone considerable stripping, and the virial radius closely resembles the stellar radius of the galaxy.  In the top panel of Figure \ref{fig:durations}, we show that more compact galaxies have shorter infall times compared to more diffuse ones, as one expects due to the nature of dynamical friction.  A more compact galaxy can plunge deeper into a larger halo before being disrupted, resulting in a MBH which is closer to the center and will  inspiral more quickly. 

 
Other orbital parameters show no apparent trends with orbital inspiral time.  For example, the duration of inspiral does not seem to have a strong dependence on the initial eccentricity or angular momentum of the BH orbit, with both circular and radial orbits having a variety of inspiral times.  The same holds true for the mass ratio of the merger, most of which take 2-4 Gyr to inspiral and merge regardless of the value of $q$.  And while compactness plays a role, the independent properties of dwarf galaxy stellar mass and galaxy radius are not correlated with inspiral time in any way on their own.

Our results are broadly consistent with those of \citet{Weller23}, who do a similar analysis but with high-redshift massive galaxies using the ASTRID simulation.   Coincidentally the galaxies in our sample are comparable in mass to theirs.  They also find fairly high initial eccentricities that decrease as the orbits progress, and overall high initial inclination angles.

\section{Evolution of the IMRIs} \label{sec:evol}


As we seek to understand the physics governing inspiral and the repercussions for gravitational wave signals, we examine how specific properties evolve with time.  Specifically, we analyze the evolution of the orbital eccentricity of the MBHs as well as the mass ratios of the system, as these quantities impact the eventual waveforms.

\begin{figure*}
\includegraphics[width=0.5\textwidth]{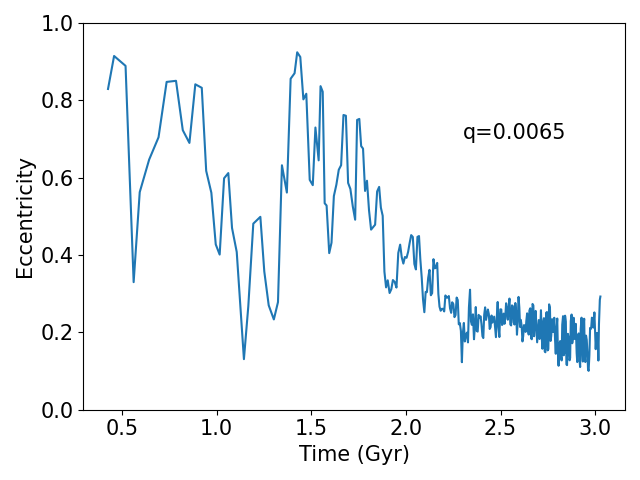}
\includegraphics[width=0.5\textwidth]{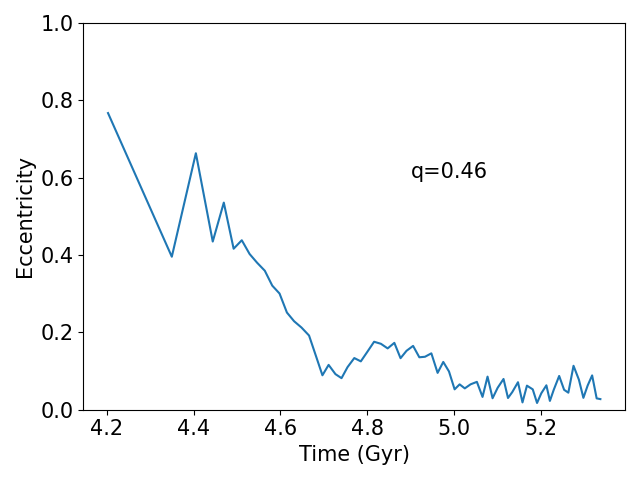}
\includegraphics[width=0.5\textwidth]{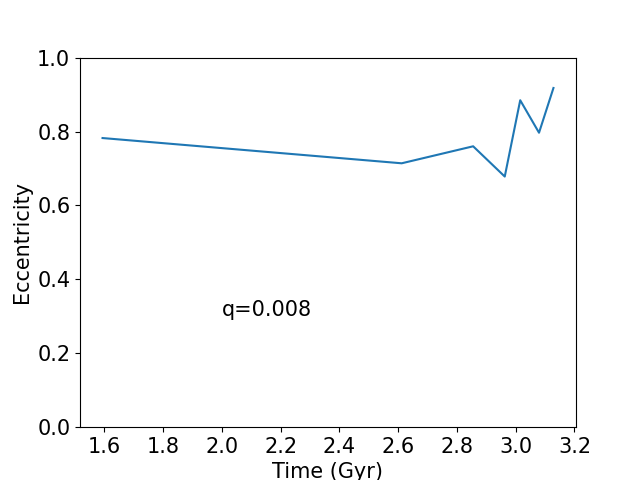}
\includegraphics[width=0.5\textwidth]{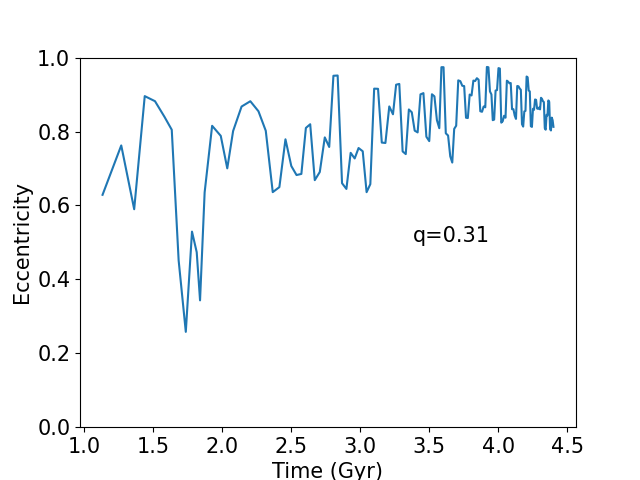}
\caption{Evolution of eccentricity with time for four sample inspirals with a range of mass ratios.  They match the panels of Figure \ref{fig:inspirals} above, though in some cases the time axis is shortened here.  In some cases the orbits become more circular with time, but in others they remain eccentric. 
\label{fig:e_evol}
}
\end{figure*}

The eccentricity of each black hole's orbit evolves after it enters the halo.  Generally speaking, each orbit becomes more circular over time.  We show examples of four such orbits in Figure \ref{fig:e_evol}, which correspond to the inspirals shown in Figure \ref{fig:inspirals}.  In some cases we truncate the time axis, because the orbital evolution is noisy at very small scales and the calculated value of $e$ is not physically realistic when $r_{\rm apo}$ is poorly defined.

\begin{figure}
\includegraphics[width=3.3in]{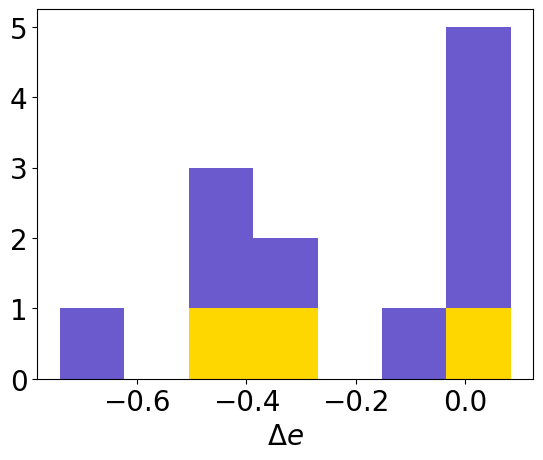}
\caption{
Change in eccentricity over the duration of the inspiral, with negative values becoming more circular.  Orbits either circularize or do not change their eccentricity.
\label{fig:deltae}
}

\end{figure}

The inspirals in the top two panels of \ref{fig:e_evol} are good examples of an initially highly eccentric orbit becoming more circular as it evolves.  This decrease in eccentricity is also seen in \citet{Weller23}.   However some orbits exhibit more chaotic behavior and do not appear to circularize -  examples are shown in the two bottom panels of the figure.  Even though the pericenter decreases over time, the eccentricity does not decrease.  In Figure \ref{fig:deltae} we show the distribution of the change in eccentricities, with negative values becoming more circular.  About half of the inspirals become more circular with time due to galactic-scale dynamical processes, while the remainder are unchanged by the dynamical influences of the inspiral.  Gravitational wave emission can be an additional circularizing effect, however, which is not accounted for here \citep{Peters64,Hinder08}; our predictions should be taken as lower limits on eccentricity evolution.

In regards to mass ratio, prior work using isolated galaxy simulations has shown that minor galaxy mergers increases accretion onto the smaller black hole substantially, such that an initial 1:10 black hole mass ratio can become 1:3 at the time of actual merging   \citep{Callegari11,Capelo15}.   However, in our simulations the mass ratios are identical between the moments in infall and coalescence.  This lack of evolution means that the infalling BH undergoes effectively no accretion as it enters the main halo.  This apparent conflict with prior results can be explained by the difference between cosmological and idealized isolated mergers;  in the latter case, the two galaxies were set up to represent a very gas-rich, $z\sim3$ Milky Way-like galaxy merging with a gas-rich dwarf.   \citet{Capelo15} analyzed 13 different versions of galaxy mergers, varying mass ratio, gas fraction, and galaxy inclination, and while some of these exhibited increased mass growth in the secondary BH, this result is not universal.  In the cosmological context all parameters are even more varied, resulting in different amounts of gas accretion.  This lack of accretion in low-mass galaxies is consistent with prior simulation results as well \citep{Bellovary19,Beckmann23}.  Minimal accretion also implies that the intrinsic spin of the BH would also not evolve as it inspirals into the center of the galaxy, which is relevant for detections by LISA.

\section{LISA repercussions}\label{sec:LISA}
 
In this work we have shown that $\sim$50\% of MBH mergers in Milky Way environments constitute IMRIs.   Mergers of these demographics should be detectable by the LISA observatory, but are less straightforward to characterize compared to major mergers (0.1 $< q < $1.0) or extreme mass ratio inspirals (EMRIs; $q < 10^{-5}$).  For gravitational wave signals to be efficiently identified, they are compared to an existing library of modeled waveforms.  Major merger waveforms are fairly straightforward to model because one can use the Post-Newtonian (PN) approximation of the numerical relativity calculation to acquire waveforms \citep{Blanchet14}.     For  EMRIs, defined as $10^{-8} < q < 10^{-5}$, waveform models are approachable because the small mass ratio lends itself to the application of a perturbative approach based on a systemic expansion of the field equations \citep{Barack19}.  However, IMRIs straddle these two regimes and neither approach is fully applicable.  Possibly one can use a combination, using perturbative and PN formalisms in domains where each are valid.

Full numerical relativity solutions are extremely computationally intensive due to the extremely high temporal and spatial resolutions required, and are difficult to compute for the vast array of IMRI parameter space \citep{Duez19,LISAWaveforms}.    Numerical methods have been used to directly calculate waveforms for mass ratios up to $q = 0.001$, but in the case of head-on black hole collisions, not quasi-circular inspirals  \citet{Lousto23}.   These simulations are a good proof of concept, but are not yet astrophysically relevant.   In the case of quasi-circular orbits, numerical relativity simulations have attained waveforms for mass ratios of 1:15 to 1:20 \citep{Jani16,SXS}, but for fewer cycles than are need for a LISA inspiral, and lacking higher-order modes.  In the event that a LISA-detected IMRI has high signal-to-noise and exhibits higher modes, there are no models of sufficient resolution to capture them \citep{Ferguson21}.  In terms of hybrid approaches, some methods may use approaches such as post-adiabatic corrections or small mass-ratio perturbation theory to reach intermediate mass ratios from the EMRI modeling regime.  However, these estimates do not include the actual merger aspect of the waveform \citep[e.g.][]{VandeMeent20,Katz21} and are therefore not useful for analyzing actual IMRI merger events. 

 The importance of highlighting this problem cannot be overstated.  While the nature of the zoom-in simulations we study here does not allow for a direct IMRI rate calculation, we can generally state that Milky Way-type galaxies are likely to host a handful of IMRI events throughout their evolution.  While $\sim$ half of these may occur at redshifts $z > 5$ and have too low signal-to-noise to be detected (see Figure \ref{fig:times}), the remainder are likely LISA sources.  Larger mass galaxies with more satellites may exhibit even more IMRIs.  Since a large fraction of mergers in complex galaxy environments may be IMRIs, the LISA community must develop the necessary tools to detect and analyze such signals.    It is likely that new approaches must be developed to generate the variety of waveforms needed to fully analyze the LISA data when it comes online \citep[e.g.][]{Rink24}.    IMRIs will add complexity to the overall signal and must be included in the global fit to properly interpret {\em all} LISA data.
 
   The black hole masses, redshifts, and frequencies of these `heavy' IMRI events can reveal much about the formation of high-redshift SMBH seeds, as well as the assembly of massive galaxies and their SMBHs.  With these signals we can determine much about the formation of the building blocks of our universe, but not if we cannot interpret LISA's observations.   An organized, concerted effort is needed to develop waveform codes and further predict the demographics of IMRI populations, so that by the mid 2030's our community will be ready for LISA's breakthroughs.
 
\section{Summary}\label{sec:concl}

We examine the four DC Justice League cosmological simulations of Milky Way-like galaxies and study the properties of black hole mergers between central supermassive black holes and the MBHs which enter the galaxy from tidally stripped dwarfs.  Our main results are as follows:

\begin{itemize}
\item About half of all dwarf-originated MBH mergers are IMRIs, with mass ratios $q < 0.04$.
\item The eccentricity of the some, but not all inspiraling MBH orbits tends to circularize with time, resulting in moderately circular orbits upon coalescence.  Mass ratios do not strongly evolve throughout the inspiral process.
\item The duration of inspiral most strongly depends on  the host galaxy compactness.
The shortest inspirals 
originate in more compact galaxies,  and the longest 
in more diffuse galaxies.
\end{itemize}

Our results are limited by the small sample size of the DC Justice League, and a broader analysis using a uniform volume simulation with realistic black hole dynamics, such as ROMULUS \citep{Tremmel17}, would more thoroughly characterize the occurrence and populations of IMRIs in Milky Way-like hosts.   Our results are also dependent on our chosen black hole seed model, which dictates not only the masses (and indirectly the mass ratios) of seeds but also where and how often seeds form.  A model with ``light seeds,'' expected to form from early massive stars, may be expected to show fewer IMRI events because dynamical friction effects scale with the mass of the orbiter and would be of lower magnitude.  However, such seeds  may be expected to form more frequently than those we study here, so one may expect an increased IMRI rate, or perhaps a cancelling out.  These uncertainties can be untangled with LISA, which can measure masses with small uncertainties.  The field must press forward with modeling waveforms along a greater parameter space in order for LISA's detections of IMRIs, and thus the complexities of seed formation, to be properly interpreted.

\begin{acknowledgments}

 We are grateful to the anonymous referee, who improved the clarity of the paper, and encouraged  expanding the discussion of the current state of waveform modeling.  JMB and YL are grateful for the support of NSF awards AST-2107764 and AST-2219090.   JMB thanks Cole Miller for productive conversations.  This research was conducted on the stolen lands of the Munsee Lenape people.
 
 \end{acknowledgments}

\vspace{5mm}
\facilities{Resources supporting this work were provided by the NASA High-End Computing (HEC) Program through the NASA Advanced Supercomputing (NAS) Division at Ames Research Center, and the Blue Waters sustained-petascale computing project.  This work also used Stampede3 at TACC  from the Advanced Cyberinfrastructure Coordination Ecosystem: Services \& Support (ACCESS) program, which is supported by U.S. National Science Foundation grants \#2138259, \#2138286, \#2138307, \#2137603, and \#2138296.}

\software{This research was analyzed with the python package {\em pynbody} \citep{pynbody}.}

\bibliographystyle{aasjournal}

\end{document}